\documentclass[5p]{elsarticle}

\usepackage{hyperref}
\usepackage{graphicx}
\usepackage{amssymb}
\usepackage{amsmath}
\usepackage{lineno}
\usepackage{color}

\biboptions{square,comma,sort&compress}

\journal{Physics Procedia}

\begin{document}

\begin{frontmatter}

\title{Structural Arrangements of Polymers Adsorbed at Nanostrings}

\author{Thomas Vogel}
\ead{thomasvogel@physast.uga.edu}
\author{Michael Bachmann}
\ead{bachmann@smsyslab.org}
\ead[url]{http://www.smsyslab.org}

\address{Soft Matter Systems Research Group, Institut f\"ur Festk\"orperforschung (IFF-2),\\ Forschungszentrum J\"ulich, 52428 J\"ulich, Germany}

\begin{abstract}
  We study ground states of a hybrid system consisting of a polymer
  and an attractive nanowire by means of computer simulations.
  Depending on structural and energetic properties of the substrate,
  we find different adsorbed polymer conformations, amongst which are
  spherical droplets attached to the wire and monolayer tubes
  surrounding it. We construct the complete con\-for\-ma\/tional phase
  diagram and analyze in more detail particularly interesting
  polymer-tube{\break} conformations.
\end{abstract}


\end{frontmatter}

\section{Introduction}
\label{intro}

The study of the interaction between organic and inorganic matter, or
in other words, the behavior of organic--inorganic systems, generates
fascinating findings with potential for novel applications in bio- and
nanotechnology.
One of the basic steps in the understanding of such systems is the
study of the adsorption of soft materials like polymers at inorganic
matter like solid substrates. In the past, numerous computational studies gave
general insights in the adsorption behavior of polymers on planar
sur\-faces~\hbox{\cite{milchev01jcp,bachmann05prl,bachmann06pre,luettmer08jcp,monika09jpcb}}.
A particularly surprising fact for example, predicted by computer
simulations and verified by experiments recently, is that a single
specific mutation in a short peptide can substantially change the
binding behavior to semiconductor
substrates~\cite{goede06lang,bachmann10acie}.

A special class of such hybrid systems are nanotubes or
nanocylinders interacting with polymers. Carbon nanotubes, for
example, are themselves quite interesting nanostructures with
surprising electronic and mechanical properties~\cite{dressel01tap},
but nanotube--polymer composites promise to enlarge the number of
possible novel applications dramatically, for example in photonics and
molecular sensor technologies~\cite{gao03ea,hasan09am}.
Theoretically, experimentally and computationally well studied is the
wetting of cylindrical substrates by liquids or polymer droplets. This
transition can be described by the crossover of barrel-like and
clamshell-like droplets~\cite{carroll86langmuir,wagner90jap,milchev02jcp}.
In another study, the adsorption behavior of individual
polymer chains on nanotubes has been studied, where a helical-like
winding of flexible and semi-flexible chains around the tubes was
found~\cite{srebnik}.\vadjust{\break}

In contrast, we will here develop a general picture of the adsorption
behavior of polymers at nanowires depending on the properties of the
substrate~\cite{vogel10prl}. For this purpose, we apply a model, where the effective
thickness and the attraction strength of the linelike substrate are variable
parameters. The above mentioned transitions and adsorbed polymer
structures are included as special cases in this picture.

\section{Model and method}
\label{model}

In our study the polymer is represented by a coarse-grained off-lattice
bead--stick model, i.e., mono\-mers do not have any inner structure
and are connected by stiff bonds. The polymer is embedded into a
three-dimensional simulation box which includes an attractive thin
string pointing into the $z$-direction. The chain is not grafted
to the string and may move freely.
The monomers interact with each other via a standard Lennard-Jones
potential
\begin{equation}
V_\mathrm{LJ}(r_{ij};\epsilon_\mathrm{m},\sigma_\mathrm{m})=4\epsilon_\mathrm{m}\,\left[\left(\frac{\sigma_\mathrm{m}}{r_{ij}}\right)^{12}-\left(\frac{\sigma_\mathrm{m}}{r_{ij}}\right)^{6}\right],
\end{equation}
where $r_{ij}$ is the geometrical distance between two monomers $i$
and $j$ and $\epsilon_\mathrm{m}$ and $\sigma_\mathrm{m}$ are set to
$1$, such that $V_\mathrm{LJ}(r_{\mathrm{min}}=2^{1/6})=-1$.
As a remnant of the origin of the model~\cite{stilli93pre} and in
order to facilitate future enhancements and the comparison with
previous studies, we introduce a weak bending stiffness, i.e., the polymer is
not flexible in a strict way, but may be considered to be flexible in
practice:
\begin{equation}
V_\mathrm{bend}(\cos\theta_i)=\kappa\,(1-\cos\theta_i)\,,
\end{equation}
where $\theta_i$ is the angle defined by the two bonds at monomer $i$
and the bending stiffness parameter $\kappa$ is here
set to $1/4$.
The interaction between monomers and the string is also based on a
simple Lennard-Jones potential, but
we neglect, as usual~\cite{milchev02jcp,monika09jpcb}, the internal structure of
the substrate, i.e., we assume a homogeneous ``charge'' distribution along
the $z$-axis. We hence simply integrate to get
\begin{align}
\nonumber
  V_\mathrm{string}(r_{\mathrm{z};i};\epsilon_\mathrm{f},\sigma_\mathrm{f})&=a\int_{-\infty}^{\infty}
  V_\mathrm{LJ}\left(\sqrt{r_{\mathrm{z};i}^2+z^2};\epsilon_\mathrm{f},\sigma_\mathrm{f}\right)\mathrm{d}z\\
&=a\,\pi\epsilon_\mathrm{f}\,\left(\frac{63\,\sigma_\mathrm{f}^{12}}{64\,r_{\mathrm{z};i}^{11}}-\frac{3\,\sigma_\mathrm{f}^{6}}{2\,r_{\mathrm{z};i}^5}\right),
\end{align}
where $r_{\mathrm{z};i}$ is the distance of the $i$th monomer perpendicular
to the string and the potential is scaled by setting $a\approx0.528$ for convenience~\cite{vogel10prl,vogel10tbp}.
$\epsilon_\mathrm{f}$ and $\sigma_\mathrm{f}$ are free
parameters and can be considered as the string attraction strength and
the effective ``thickness'' of the string, which is proportional to
the equilibrium distance of the string potential, respectively.
The overall energy of the system finally reads
\begin{equation}
E=\sum_{i=1,j>i+1}^{N-2}V_\mathrm{LJ}(r_{ij})+\sum_{i=2}^{N-1}V_\mathrm{bend}(\cos\theta_i)+\sum_{i=1}^N V_\mathrm{string}(r_{\mathrm{z};i})\,.
\end{equation}
For estimating the ground-state energies, we apply
gen\-er\-al\-ized-ensemble Monte Carlo
methods~\cite{bergneuh92prl,wangl01prl}. Conformational changes are
forced by applying a variety of update moves, including local
crankshaft and slithering-snake moves and global spherical-cap and
translation moves~\cite{vogel10tbp}.

\section{Results}
\label{results}

\begin{figure}[b!]
\includegraphics[width=\columnwidth,clip]{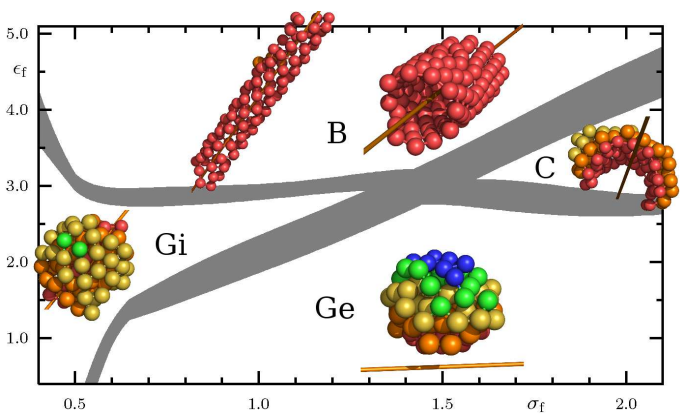}
\vspace{-5mm}
\caption{The low-energy conformational phase diagram for polymers
  adsorbed at nanostrings. From bottom to top, the string attraction
  strength $\epsilon_\mathrm{f}$ increases, from left to right, the effective radius of the
  string $\sigma_\mathrm{f}$ becomes larger. Different monomer colors or shadings encode
  different distances from the string. Monomers near the equilibrium
  distance from the string (colored in red) are defined to be in
  contact with it.}
\label{fig1}
\end{figure}

We now discuss low-energy structures of the above
described system for different values of the string-potential
parameters $\sigma_\mathrm{f}$ and $\epsilon_\mathrm{f}$. Based on the
simulation of more than hundred system parametrizations, we construct the full
conformational phase-diagram which is shown in Fig.~\ref{fig1}.

\begin{figure}
\includegraphics[width=\columnwidth,clip]{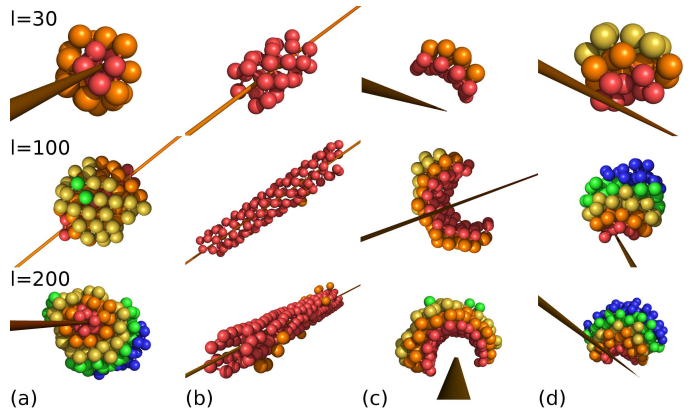}
\vspace{-5mm}
\caption{Visualizations of low-energy conformations with $N=30$, $100$,
  and $200$ monomers in phases (a) Gi, (b) B,\break (c) C, and (d) Ge.}
\label{fig2}
\end{figure}

We identify four major conformational phases, which we denote Gi,
B, C, and Ge. For small values of both $\sigma_\mathrm{f}$ and
$\epsilon_\mathrm{f}$, i.e., for weak string attraction and small
effective radius of the string, we find globular conformations with
spherical symmetry surrounding the string (phase Gi). Increasing the
string attraction strength, conformations stretch out along the string
breaking the spherical symmetry and barrel-like conformations with the
string inside emerge (phase B). In the case of very high string
attraction we even find monolayer tubes with each monomer being in
direct contact with the substrate. Due to the finite size of the
system, these barrel-like structures break when increasing the
effective diameter of the string and low-energy structures become
clamshell-like (phase C), i.e., we find adsorbed conformations
consisting of a few layers which are not wrapping the string
completely. Finally, decreasing the string attraction at this
effective radius, conformations become spherical droplets sticked to
the string (phase Ge). Low-energy conformations from different regions
are visualized exemplarily in Figs.~\ref{fig1} and~\ref{fig2}.\break We
convinced ourselves by simulating chains with lengths $N=30$ and $200$
that the general, qualitative structure of the conformational phase
diagram does not depend on the actual length of the polymer, see
Fig.~\ref{fig2} for examples. Of course, details like the exact
positions of transitions lines may indeed depend on the actual polymer
length.

\begin{figure}[t]
\begin{center}
\includegraphics[width=.8\columnwidth]{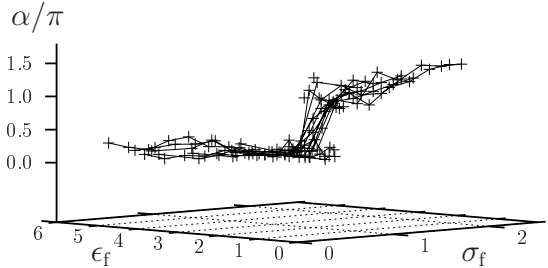}
\end{center}
\vspace{-3mm}
\caption{Opening angle, data points with same $\sigma$-value are
  connected by lines to guide the eyes. See also Fig.~\ref{fig2}(a) and
  (b) for closed conformations ($\alpha/\pi\approx0$) and
  \ref{fig2}(c) and (d) for open ones ($\alpha/\pi>1$).}\vspace{3mm}
\label{fig3}
\end{figure}
\begin{figure}[b!]
\begin{minipage}[b]{\columnwidth}
\includegraphics[width=\textwidth]{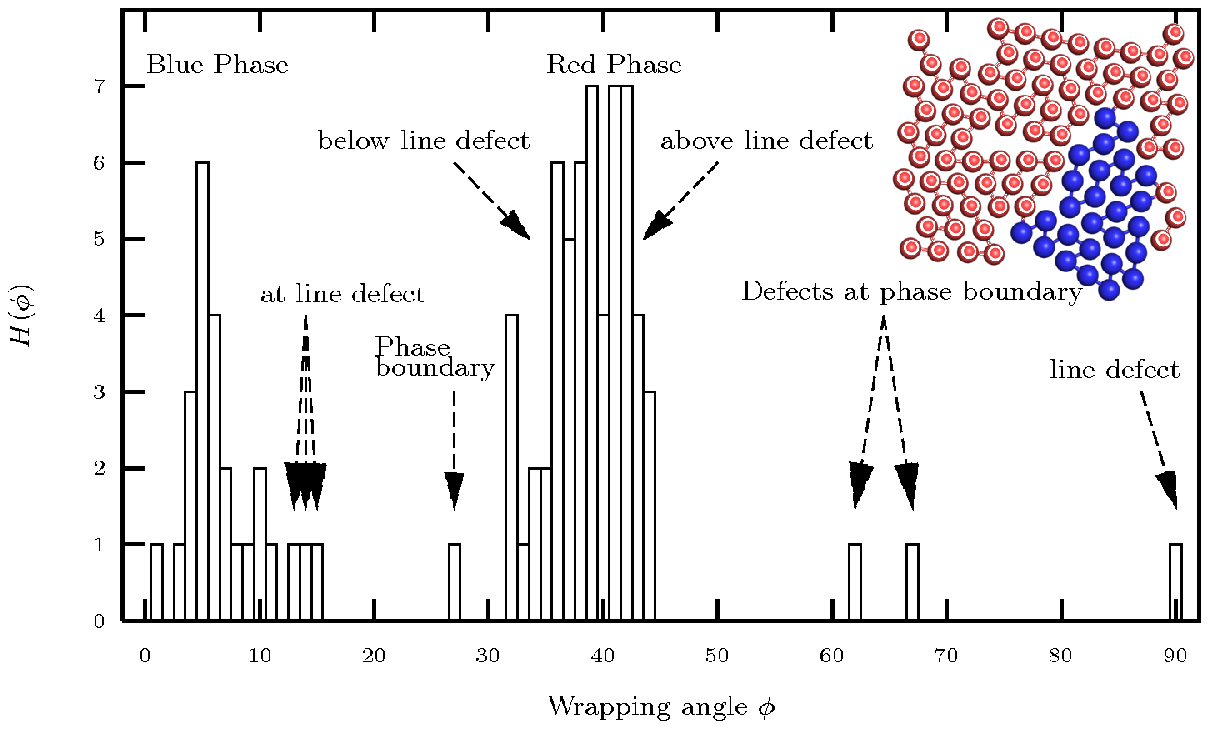}\\
\includegraphics[width=.4\textwidth]{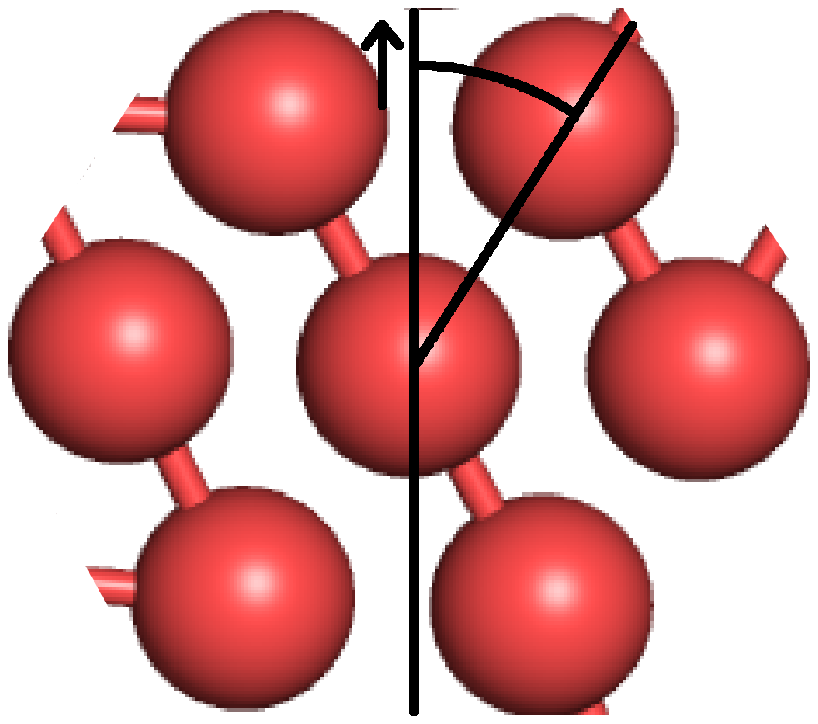}
\end{minipage}
\begin{picture}(0,0)
\put(52,73){$\phi_i$}
\put(45,56){$i$}
\put(0,103){a)}
\put(0,13){b)}
\end{picture}
\vspace{-3mm}
\caption{a) Angular distribution function of a monolayer structure
  with $\sigma_{\mathrm{f}}=1.569$ (also shown as inset picture in
  Fig.~\ref{fig1}, phase B). Inset: Unzipped, planar
  representation. Different colors (red and blue, or light- and
  dark-gray, respectively) represent main regions with different wrapping angles. b)
  Illustration of the definition of the wrapping angle. The arrow
  points in the direction of the string.}
\label{fig4}
\end{figure}

To define the different phases and the crossover between them, we
introduce observables showing a peculiar behavior at the
transitions. In Gi and B, for example, the polymer conformations
surrounds the string completely, in contrast to the structures in Ge
and C. For the localization for the transition between
Gi$\leftrightarrow$Ge and B$\leftrightarrow$C, we hence look at the
opening angle~$\alpha$~\cite{vogel10prl} of the polymer conformation.
The value of this angle shows a jump (low values for Gi and B, high
values for C and Ge) at the crossover between these phases, which is
shown in Fig~\ref{fig3}. See~\cite{vogel10prl} for more details on the
localization of phase boundaries.

We finally would like to get at a deeper analysis of some structures
in phase B. At very high attraction strengths, low-energy
conformations become regular monolayer conformations wrapped around
the string, i.e., single-walled tubes with an ordered arrangement of
monomers form. It is noticeable, that there is a competition between
different chiral angles, i.e., orientations of the wrapping. This
behavior is of particular interest as it is in a similar manner known
from carbon nanotubes~\cite{dressel01tap}. Figure~\ref{fig4}
illustrates the distribution of the chiral or wrapping angles of such
a monolayer conformation. Therefore we unzip the structure, i.e., we
project it onto a~plane, and measure the angular distribution function
(adf) of this unzipped structure, whereas we define the chiral or
wrapping angle $\phi_i$ of the $i$th monomer as the smallest angle
between the vectors pointing to its neighbors and the vector in the
string direction (see Fig.~\ref{fig4}\,b). In the adf we clearly see
the signals from two different regions with different chiralities as
well as from defects in that structure. A detailed, systematic
analysis of monolayer structures at high string attraction strength
and different effective string thicknesses is subject of ongoing
studies~\cite{vbma_tbp1,vbma_tbp2}.

\medskip

\paragraph{Acknowledgment:} 
We would like to thank J. Adler and T. Mutat from the Technion Haifa
for intense discussions. This work is supported by the Umbrella
program under Grant No. SIM6 and by supercomputer time provided by the
FZ J\"ulich under Project Nos. jiff39 and jiff43.

\end{document}